\documentclass{article}
\usepackage{amsmath,graphicx,mlspconf}
\usepackage{graphicx}
\copyrightnotice{979-8-3503-2411-2/23/\$31.00 {\copyright}2023 IEEE}

\toappear{2023 IEEE International Workshop on Machine Learning for Signal Processing, Sept.\ 17--20, 2023, Rome, Italy}

\title{Semantic-Aware Image Compressed Sensing}
\name{%
    Bowen Zhang$^{\star}$
    \qquad Zhijin Qin$^{\dagger}$ \thanks{The work was supported in part by the National Natural Science Foundation
of China (NSFC Nos 62293484) and the DCMS Future Open Networks Research Challenge Programme of UK (PA5657/TUDOR). (Corresponding author: Zhijin Qin.)}
    \qquad Geoffrey Ye Li$^{\star}$\thanks{The emails of the authors are \{k.zhang21, geoffrey.li\}@imperial.ac.uk, qinzhijin@tsinghua.edu.cn.}%
}
\address{
    $^{\star}$ Department of Electrical and Electronic Engineering, Imperial College London, London, UK \\%
    $^{\dagger}$ Department of Electronic Engineering, Tsinghua University, Beijing, China%
}

\begin{document}
\ninept

\maketitle

\begin{abstract}
Deep learning based image compressed sensing (CS) has achieved great success. However, existing CS systems mainly adopt a fixed measurement matrix to images, ignoring the fact the optimal measurement numbers and bases are different for different images. To further improve the sensing efficiency, we propose a novel semantic-aware image CS system. In our system, the encoder first uses a fixed number of base CS measurements to sense different images. According to the base CS results, the encoder then employs a policy network to analyze the semantic information in images and determines the measurement matrix for different image areas. At the decoder side, a semantic-aware initial reconstruction network is developed to deal with the changes of measurement matrices used at the encoder. A rate-distortion training loss is further introduced to dynamically adjust the average compression ratio for the semantic-aware CS system and the policy network is trained jointly with the encoder and the decoder in an end-to-end manner by using some proxy functions. Numerical results show that the proposed semantic-aware image CS system is
superior to the traditional ones with fixed measurement matrices.
\end{abstract}
\section{Introduction}
\label{sec:intro}

The traditional image acquisition systems based on the Nyquist-Shannon sampling theorem require the sampling ratio of image sensors to be no less than twice the bandwidth of the original signal \cite{shannon1949communication}, which is unfriendly to the applications where inexpensive sensors shall be used or oversampling may be harmful to the object being sensed (e.g. medical imaging). Also, as many sensed images will be compressed for storage or transmission purposes, the sensing costs for pixels that will be discarded in the compressing process are higher than needed in traditional sensors. Based on these considerations, image compressive sensing (CS) that jointly implements the sampling and compression processes has been proposed as a new paradigm for image acquisition and reconstruction \cite{duarte2008single, oike2012cmos}. The CS theory \cite{donoho2006compressed} also shows the number of measurements required for image CS is much fewer than suggested by the Nyquist-Shannon sampling rate as images can be well sparsely represented. 

Recently, deep learning based image CS methods have been developed to improve the sensing efficiency and reconstruction accuracy in image CS problem. For example, based on the block-based image CS architecture \cite{gan2007block,mun2009block}, Shi \textit{et} \textit{al}. \cite{shi2019image} propose a convolutional neural network (CNN)-based image CS network architecture, CSNet, where sensing matrices and reconstruction network are jointly optimized.  Motivated by the iterative algorithm, deep unfolding networks, such as ADMM-Net \cite{sun2016deep} and AMP-Net \cite{zhang2020amp}, are introduced as reconstruction networks for image CS, which balances reconstruction speed and network interpretation. To address the problem of CNN-based networks in modelling long distance relationships, a cascaded visual transformer (ViT) architecture is developed in \cite{lorenzana2022transformer}. The information bottleneck measurement in \cite{lee2022information} can enhance the training process of sensing network by explicitly modelling the importance level of different measurements. 

Despite the fast development of image CS methods, existing methods mainly use a fixed sensing measurement matrix for different images.
Recent research on semantic communications has, however, demonstrated that data transmission efficiency can be increased if the communication policy is modified in accordance with the semantic information in the data \cite{qin2021semantic,xie2021deep,zhang2023semantic}. This inspires us to think if data acquisition process can also be improved in a semantic-aware manner. In fact, instead of using a fixed sensing matrix, images with varying types of semantic information shall be sensed and compressed by different measurement matrices, including different numbers of measurements and different measurement bases\footnote{Each row of the measurement matrix is called as a measurement base in this paper}. It is well-known that different semantic information will have different sparsity levels when they are represented sparsely under a sparse transformation matrix. From CS theory, more measurements shall be used for signals less sparse to satisfy the restricted isometric property (RIP) requirement \cite{candes2008introduction,candes2011compressed}. Therefore, the sparse signal cannot be well recovered if fewer samples than required are collected; but the sensing costs are higher than needed if more samples are used. This inspires us to adjust the number of measurements according to the semantic information type. 

Furthermore, it is also helpful to adjust the measurement bases for different semantic information for image CS problem. Specifically, the sparse representations of different semantic information may have different support sets (or sparsity patterns)\footnote{considering the case where different semantic information has different frequency components and a discrete Fourier transform (DFT) matrix is used as the sparse transformation matrix.}. Without these support information, one need to ensure that the correlations between all pairs of columns of measurement matrix are small enough so that the sparse recovery methods, such as orthogonal matching pursuit \cite{tropp2007signal}, can operate successfully. By contrast, if the support set information can be roughly estimated by analysing the semantic information type and is available before CS process, we can enhance the sensing and sparse signal recovery process by explicitly reducing the correlations of the columns belonging to the support set and reducing the search space in recovery process. 

There are two challenges for this semantic-aware CS process: 1) how to estimate the semantic information of an area and use it to adjust the sensing process before sensing it; 2) how to dynamically adjust the measurement matrix according to the semantic information type without storing a matrix per type. To address the first challenge, we divide the CS process into two steps. In the first step, we use a fixed measurement matrix for all areas and estimate the semantic information from these observations for each area. The estimated semantic information is then used to decide the measurement matrix for different areas in the second step. To deal with the second challenge, we learn a relatively large measurement matrix and dynamically select rows from this large matrix to construct the semantic-aware measurement matrix for each individual area. The selection process is done by a policy network, which is trained jointly with the measurement matrix and the reconstruction network via some proxy functions in an end-to-end manner. Note that the whole network follows the designs in block-based image CS problem (BCS)\cite{gan2007block, mun2009block, shi2019image}.

The most related work to ours is the content-aware scalable network (CASNet) proposed in \cite{chen2022content}. Our work differs from \cite{chen2022content} in the following three aspects: 1) Our method adjusts both the number of measurements and the measurement bases; 2) Instead of using the same compression ratio from different images, we adjust the compression ratio for different images under the constraint that the average compression ratio over the the training/validation data-set meets the requirement. 3) To reduce memory and computational costs for sensor, our policy network works on the measurement space, which is far smaller than the image signal space.  

\section{Problem formulation and system model}
In this section, we will introduce the problem formulation and system models for existing network-based BCS and the proposed semantic-aware BCS.
\subsection{Block-based image compressive sensing}
Given an image $I\in \mathcal{R}^{H\times W\times3}$, BCS first divides the image into non-overlapping
blocks of size $B\times B\times3$ and reshapes the blocks into vectors. Then, each block is sensed by a learned measurement matrix $\mathcal{\phi}$ of size $n\times 3B^{2}$. This process can be represented as,
\begin{equation}
y_{i,j}=\mathcal{\phi}x_{i,j},
\label{equ:1}
\end{equation}
for $i=1,2,\cdots,\frac{H}{B}$, $j=1,2,\cdots,\frac{W}{B}$, where $x_{i,j}$ and $y_{i,j}$ are the $(i,j)$-th block in 2D image space and corresponding measurements. In this process, the compression ratios for each block and the whole image are the same, \textit{i}. \textit{e}., $\frac{n}{3B^{2}}$. In this work, we set $H=W=224$ and $B=32$, resulting in $7\times 7$ blocks.

After sensing, the goal of BCS is to reconstruct the original image from these CS measurements. In this work, we mainly focus on network-based BCS methods \cite{shi2019image}. In particular, after obtaining CS measurements, network-based BCS first obtains an initial reconstructed image via a trainable matrix $\mathcal{\theta}$ of size $3B^{2}\times n$ \cite{shi2019image}. Given CS measurement $y_{i,j}$ of the $(i,j)$-th block, its initial reconstruction result $\hat{x}_{i,j}$ can be represented as,
\begin{equation}
\hat{x}_{i,j}=\mathcal{\theta}y_{i,j}.
\label{equ:2}
\end{equation} 
To this end, the initial reconstruction result for each block is still a vector. Network-based BCS methods will further reshape and concatenate these reconstructed vectors to get an initial reconstructed image $\hat{I}$ \cite{shi2019image}.

After initial reconstruction, a deep reconstruction network $D(\cdot)$ is utilized to refine the initial reconstruction result,
\begin{equation}
\tilde{I}=D(\hat{I}),
\label{equ:3}
\end{equation}
where $\tilde{I}$ denotes the final reconstructed images. Depending on the network architecture, the deep reconstruction network can be categorized as model-driven networks \cite{sun2016deep,zhang2020amp}, data-driven networks \cite{shi2019image,lorenzana2022transformer}, and hybrid networks \cite{chen2022content}. 
\begin{figure}[t]
\centering
\includegraphics[scale=0.53]{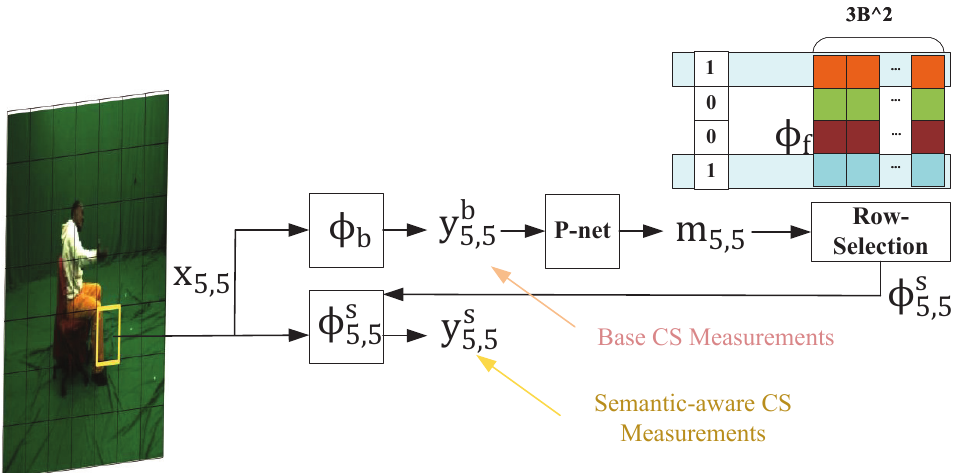}
\caption{The basic idea of the proposed semantic-aware block-based image compressed sensing system.}
\label{fig:adaptive_image_overview}
\end{figure}

\subsection{Semantic-aware block-based image compressive sensing}
Based on the BCS methods, we now give the pipeline of the proposed semantic-aware BCS methods, which is also shown in Fig. \ref{fig:adaptive_image_overview}. As aforementioned, the semantic-aware BCS is divided into two steps. In the first step, a learned base measurement matrix $\mathcal{\phi}_{b}$ of size $n_{b}\times 3B^{2}$ is utilized to sense each block as follows,
\begin{equation}
y_{i,j}^{b}=\mathcal{\phi}_{b}x_{i,j},
\label{equ:4}
\end{equation}
where $y_{i,j}^{b}$ is the CS measurements under base measurement matrix for the $(i,j)$-th block. 

After obtaining these base CS measurements, a policy network $P(\cdot)$ will take these measurements as inputs, analyse its semantic information type, and generate 0-1 row-selection vectors for each block, which can be represented as,
\begin{equation}
m_{1,1},m_{1,2},\cdots,m_{\lceil \frac{H}{B} \rceil,\lceil \frac{W}{B} \rceil} =P(y_{1,1}^{b},y_{1,2}^{b},\cdots, y_{\lceil \frac{H}{B} \rceil, \lceil \frac{W}{B} \rceil}^{b}),
\label{equ:5}
\end{equation}
where $m_{i,j}\in \{0,1\}^{n_{max}}$ and $n_{max}$ is the number of rows of a shared large measurement matrix $\mathcal{\phi}_{f} \in \mathcal{R}^{n_{max}\times 3B^{2}}$. The sensor then constructs semantic-aware measurement matrix $\mathcal{\phi}_{i,j}^{s}\in \mathcal{R}^{n_{i,j}^{s}\times 3B^{2}}$ by selecting rows from $\mathcal{\phi}_{f}$ according to the locations of values $1$ in $m_{i,j}$ for each block, where $n_{i,j}^{s}$ is the number of values $1$ in $m_{i,j}$. Usually, $n_{i,j}^{s}$ will have a higher value for blocks where the sparse representations are less sparse. 

Next, each $\mathcal{\phi}_{i,j}^{s}$ is utilized to sense the corresponding area in the second step. This process can be represented as,
\begin{equation}
y_{i,j}^{s}=\mathcal{\phi}_{i,j}^{s}x_{i,j},
\label{equ:7}
\end{equation}
where $y_{i,j}^{s}$ is the CS measurements under the learned semantic-aware measurement matrix for the $(i,j)$-th block. After these two steps, there are $n_{b}+n_{i,j}^{s}$ measurements for the $(i,j)$-th block. The average compression ratio $r_{avg}$ for an image dataset with $N$ images can be calculated as $r_{avg}=\frac{n_{avg}}{3B^{2}}$, where $n_{avg}=\frac{1}{N\lceil \frac{HW}{B^{2}} \rceil}\sum_{k=1}^{N}\sum_{i=1}^{\lceil \frac{H}{B^{2}} \rceil}\sum_{j=1}^{\lceil \frac{W}{B^{2}} \rceil} (n_{i,j}^{s,k}+n_{b})$ denotes the average number of measurements per block, where $n_{i,j}^{s,k}$ is the number of measurements for the $(i,j)$-th block in image $k$ at the second step.

Following BCS, semantic-aware BCS also has initial reconstruction and deep reconstruction stages. However, different from BCS, semantic-aware BCS needs to tackle the changes of measurement matrices used in the sensing stage. Specifically, since $y_{i,j}^{s}$ is generated through different $\mathcal{\phi}_{i,j}^{s}$ in each block, it is hard to reconstruct initial reconstruction result $\hat{x}_{i,j}$ through a shared matrix $\theta$ for all blocks. Therefore, in the initial reconstruction stage, we first generate a block-wise matrix $\theta_{i,j}^{s}$ for each block using another weight-generation network $A(\cdot)$, which takes base measurements obtained from $\mathcal{\phi}_{b}$ as inputs,
\begin{equation}
\mathcal{\theta}_{1,1}^{s},\mathcal{\theta}_{1,2}^{s},\cdots, \mathcal{\theta}_{\lceil \frac{H}{B} \rceil, \lceil \frac{W}{B} \rceil}^{s}=A(y_{1,1}^{b},y_{1,2}^{b},\cdots, y_{\lceil \frac{H}{B^{2}} \rceil, \lceil \frac{W}{B^{2}} \rceil}^{b}),
\label{equ:8}
\end{equation}
where $\mathcal{\theta}_{i,j}^{s} \in \mathcal{R}^{3B^{2}\times n_{i,j}^{s}}$ is the generated initial reconstruction matrix for block $(i,j)$. After that, $\hat{x}_{i,j}$ can be represented as,
\begin{equation}
\hat{x}_{i,j}=\left[ \mathcal{\theta}_{b}, \mathcal{\theta}_{i,j}^{s}\right]\left[ \begin{array}{c}
y_{i,j}^{b}  \\
y_{i,j}^{s}  \end{array} \right],
\label{equ:9}
\end{equation}
where $\mathcal{\theta}_{b} \in \mathcal{R}^{3B^{2}\times n_{b}}$ is the initial reconstruction matrix for base measurements. In this work, we do not directly learn the $\mathcal{\theta}_{i,j}^{s}$ due to the large matrix size. We decompose $\mathcal{\theta}_{i,j}^{s}$ into a large matrix $\mathcal{\theta}_{s} \in \mathcal{R}^{3B^{2}\times n_{max}}$, which is shared among blocks, and a small matrix $\mathcal{\tilde{\theta}}_{i,j}^{s} \in \mathcal{R}^{n_{max}\times n_{i,j}^{s}}$, which is actually learned for each block. Here, $n_{max}(\ll 3B^{2})$ is a pre-defined value representing the maximum number of measurements for each block in step two. With this decomposition, Eq. (\ref{equ:9}) is re-written as,
\begin{equation}
\hat{x}_{i,j}=\left[ \mathcal{\theta}_{b}, \mathcal{\theta}_{s}\right]\left[ \begin{array}{cc}
I_{n_{b}\times n_{b}}&0  \\
0& \mathcal{\tilde{\theta}}_{i,j}^{s} \end{array} \right]\left[ \begin{array}{c}
y_{i,j}^{b}  \\
y_{i,j}^{s}  \end{array} \right],
\label{equ:10}
\end{equation}
The deep reconstruction network should also be designed in a way adaptive to the changes of $\mathcal{\phi}_{i,j}^{s}$ in different blocks, which will be considered in our future work. In this work, we use a memory-friendly deep reconstruction network for simplicity. More details will be given hereafter. 

If we substitute semantic-aware block-wise matrices $\mathcal{\phi}_{i,j}^{s}$ in Eq.(\ref{equ:7}) with a fixed matrix $\mathcal{\phi}_{bf}$ of size $(n_{avg}-n_{b})\times 3B^{2}$ and $\mathcal{\theta}_{i,j}^{s}$ in Eq.(\ref{equ:9}) with a fixed matrix $\mathcal{\theta}_{bf}$ of size $3B^{2}\times(n_{avg}-n_{b})$ for different blocks in the above architecture, the semantic-aware BCS can easily degrade to the traditional BCS with the same average compression ratio, enabling a fair comparison between semantic-aware BCS and traditional BCS. 
\begin{figure*}[ht]
\centering
\includegraphics[scale=0.65]{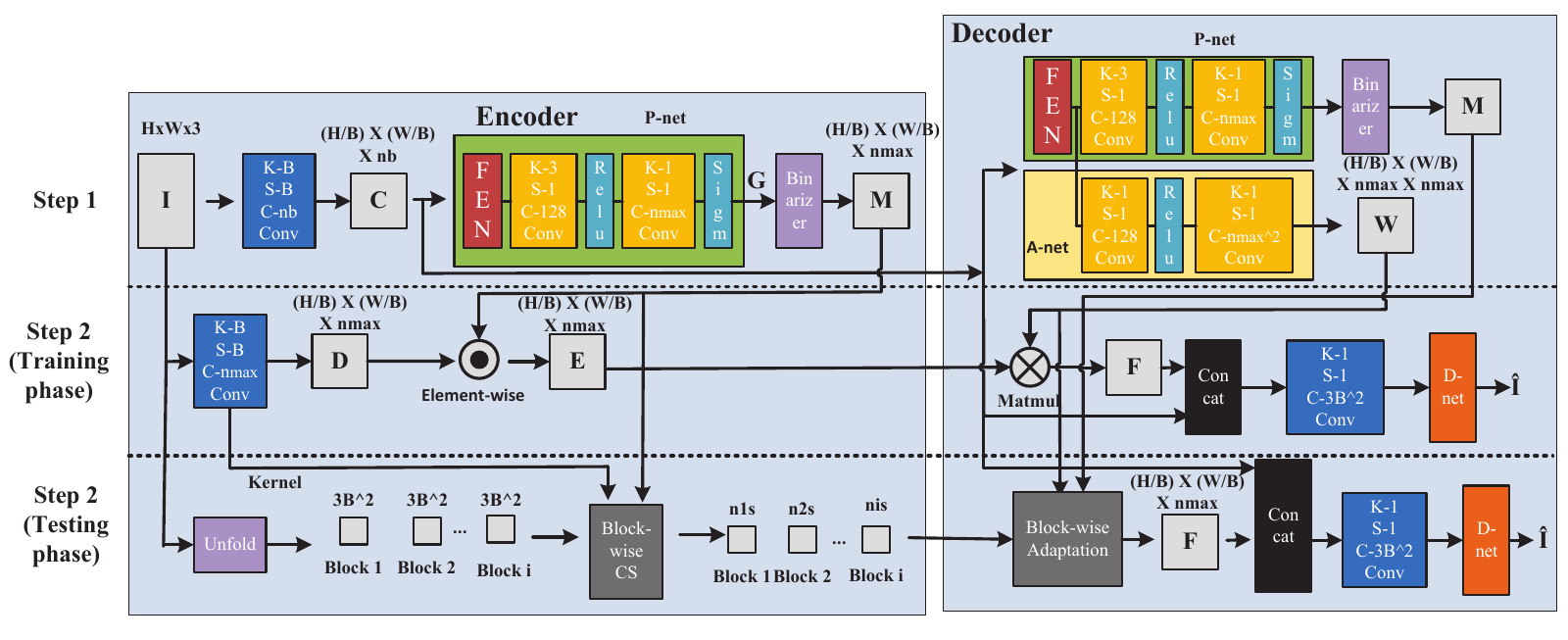}
\caption{The architecture of the proposed semantic-aware block-based image compressed sensing system. In the figure, K-A, S-B, C-n Conv denotes a convolution layer with $A\times A $ kernels, stride $B$, and $n$ output channels. The convolution layers with blue color are trained without bias parameters while the convolution layers with yellow color have bias parameters.}
\label{fig:adaptive_image}
\end{figure*}

\section{Semantic-aware image BCS network}
In this section, we will introduce the network architectures of the proposed semantic-aware image BCS network and the training details.

\subsection{Network architecture}
As shown in Fig. \ref{fig:adaptive_image}, the architecture of the proposed coding method is composed of an encoder and a decoder. We first introduce the encoder. Given an image $I$ of size $H\times W \times 3$, we first apply a $B\times B$ convolution layer (Conv) with a stride size of $B$ and $n_{b}$ output channels to $I$, generating features $C \in \mathcal{R}^{\frac{H}{B}\times \frac{W}{B}\times n_{b}}$. This process corresponds to Eq.(\ref{equ:4})\footnote{More details for this step are explained in \cite{shi2019image}} and $c_{ij}=C\left[i,j,:\right]\in \mathcal{R}^{n_{b}}$, for $i=1,2,\cdots,\frac{H}{B}$ and $j=1,2,\cdots,\frac{W}{B}$, denotes the base CS measurements for the image area $I[(i-1)B:iB,(j-1)B:jB,:]$, which is also called as $(i,j)$-th block in previous section. 

Next, a policy network, denoted as P-net, takes $C$ as inputs and generates an intermediate feature $G\in \mathcal{R}^{\frac{H}{B}\times \frac{W}{B}\times n_{max}}$, which is then quantified into a 0-1 mask matrix $M$ by a binarizer. The detailed architecture of P-net is shown in Fig. \ref{fig:adaptive_image}, where FEN denotes a feature extraction network consisting of three $3\times 3$ Convs with stride $1$ and $256$ output channels, and $sigm$ denotes sigmoid activation layer. The binarizer is used to conduct binary quantization to $G$. It outputs $1$ if the input is over $0.5$; otherwise it outputs $0$. Due to binarizer, the backward gradient is zero almost everywhere, restricting the parameter update of the P-net. To solve this non-differentiate issue, we use a straight-through estimator of the gradient \cite{courbariaux2016binarized} which directly uses the gradients to $M$ as the gradients to $G$. The generation of $m_{i,j}=M[i,j,:]\in \{0,1\}^{n_{max}}$ corresponds to Eq.(\ref{equ:5}).

After that, we apply a $B\times B$ Conv layer with stride size $B$ and $n_{max}$ output channels to $I$ and obtain features $D \in \mathcal{R}^{\frac{H}{B}\times \frac{W}{B}\times n_{max}}$. This process is equivalent to sense each block using the shared large measurement matrix $\mathcal{\phi}_{f}$ mentioned before. We then multiple $D$ and 0-1 mask matrix $M$ element-wisely to get $E$. $e_{i,j}=E[i,j,:]\in \mathcal{R}^{n_{max}}$ can be regarded as a zero-padded version of $y_{i,j}^{s}$ in Eq.(\ref{equ:7}), where the values of unimportant measurements indicating by the values $0$ in $m_{i,j}$ are set as $0$. 

Note that in the training process, we first sense each block with the maximum number of measurements and then set unimportant ones to $0$. In this way, the number of measurements generated by each block is the same, making it easier to implement the batch training method. However, during the test phase, we first need to construct $\mathcal{\phi}_{i,j}^{s}$ by considering $m_{i,j}$ and $\mathcal{\phi}_{f}$ stored in the Conv layer and then sense each block with $\mathcal{\phi}_{i,j}^{s}$ to get $y_{i,j}^{s}$. Only in this way can we reduce the actual sensing costs. The differences between the training phase and the testing phase are shown in Fig. \ref{fig:adaptive_image}.

At the decoder side, base CS measurements $C$ are first fed into the P-net, which shares the same architecture and parameters as the one in the encoder. And then, the outputs of FEN in the P-net are used as the inputs of a weight-generation network, A-net, which generates weights $W \in \mathcal{R}^{\frac{H}{B}\times \frac{W}{B}\times (n_{max}\times n_{max})}$. Here, $w_{i,j}=W[i,j,:]\in \mathcal{R}^{n_{max}\times n_{max}}$ is similar to the $\mathcal{\tilde{\theta}}_{i,j}^{s}$ in Eq.(\ref{equ:10}). The only difference is that more columns are generated in $w_{i,j}$ than $\mathcal{\tilde{\theta}}_{i,j}^{s}$ for zero-padded values in $e_{i,j}$. The redundant columns will not affect the final results. In the training phase, for block $(i,j)$, we conduct matrix multiplication between $w_{ij}$ and $e_{ij}$, and repeat this process for all blocks. This process corresponds to operation $\mathcal{\tilde{\theta}}_{is}y_{is}$ in Eq.(\ref{equ:10}). The results are features $F\in \mathcal{R}^{\frac{H}{B}\times \frac{W}{B}\times n_{max}}$. After obtaining $F$, we concatenate $F$ and $C$ along the channel dimension, which equals forming the $[Iy_{ib};\mathcal{\tilde{\theta}}_{is}y_{is}]$ in Eq.(\ref{equ:10}). Following that, we input them into a $1\times 1$ Conv with stride $1$ and $3B^{2}$ output channels. The convolution operation represents multiplying $[\mathcal{\theta}_{b}, \mathcal{\theta}_{s}]$ by the left in Eq.(\ref{equ:10}). In the testing phase, operations are slightly different. We first select columns from $W$ under the guidance of $M$ and obtain $\mathcal{\tilde{\theta}}_{i,j}^{s}$, which is then used to multiply with $y_{i,j}^{s}$. 

\begin{table}[]
\centering
\caption{Detailed architecture of D-net.}
\label{tab:my-table}
\scalebox{0.9}{
\begin{tabular}{|c|c|c|c|}
\hline
Layer & Type           & Parameters          & Output\_shape                        \\ \hline
0     & Input          & -              & (H/B)x(W/B)x3$B^{2}$ \\ \hline
1     & Depth-to-Space & block-size=B/8 & (H/8)x(W/8)x192                   \\ \hline
2     & Conv           & K-3,S-1,C-256  & (H/8)x(W/8)x256                   \\ \hline
3     & Resblock       & K-3,S-1,C-256  & (H/8)x(W/8)x256                   \\ \hline
4     & Conv           & K-3,S-1,C-192  & (H/8)x(W/8)x192                   \\ \hline
5     & Depth-to-Space & block-size=2   & (H/4)x(W/4)x48                    \\ \hline
6     & Conv           & K-3,S-1,C-128  & (H/4)x(W/4)x128                   \\ \hline
7     & Resblock       & K-3,S-1,C-128  & (H/4)x(W/4)x128                   \\ \hline
8     & Conv           & K-3,S-1,C-48   & (H/4)x(W/4)x48                    \\ \hline
9     & Depth-to-Space & block-size=4   & HxWx3                             \\ \hline
10    & Conv           & K-3,S-1,C-64   & HxWx3                             \\ \hline
11    & Resblock       & K-3,S-1,C-64   & HxWx3                             \\ \hline
12    & Conv           & K-3,S-1,C-3    & HxWx3                             \\ \hline
\end{tabular}}
\end{table}
 At last, a D-net is used for deep reconstruction. We show the architecture of the D-net in Table \ref{tab:my-table}, where Depth-to-Space layer is used to rearrange features from the channel dimension into spatial dimension and Resblock layer is a cascaded of Conv-relu-Conv with skip connections. The parameters of the Convs used in Resblock are shown in Table \ref{tab:my-table}. Note that as the main goal of this work is to verify the semantic-aware operations introduced in the encoder and the initial reconstruction process in the decoder, we do not spend much effort in designing the deep reconstruction network, whose designs, however, are important for comparison with state-of-the-arts BCS works and will be left for our future work. In this work, the adopted D-net has comparable performance with the one used in CSNet \cite{shi2019image}. Different from CSNet whose D-net operates on image signal space and thus has a high memory and computational costs, the D-net shown in Table \ref{tab:my-table} extracts features mainly in feature space with low spatial dimension and has faster speed and lower memory consumption.
\subsection{Rate-distortion trade-off}
Here, we will describe the training process of the whole network. In this semantic-aware BCS, we hope the number of CS measurements to be small while at the same time the peak signal-to-noise ratio (PSNR) to be high. Under this design goal, we can formulate the training loss as the well-known rate-distortion trade-off \cite{zhang2022semantic}, which can be defined as follows,
\begin{equation}
\mathcal{L}=\sum_{I\in \mathcal{I}}{\mathcal{L}_{2}(I,\hat{I})+\gamma \mathcal{L}_{R}(I)},
\label{equ:11}
\end{equation}
where $\mathcal{I}$ denotes the training dataset, $\mathcal{L}_{2}=||I-\hat{I}||_{2}^{2}$ denotes the distortion, and $\mathcal{L}_{R}(I)=\sum_{i=1}^{H/B}\sum_{j=1}^{W/B} \sum_{k=1}^{n_{n_{max}}} G(I)[i,j,k]$ denotes the rate loss, and $G(I)$ is the outputs of P-net when $I$ is the network input. As $G(I)$ determines the number of values $1$ in $M$, minimizing $G(I)$ equals minimizing the number of measurements. Besides, $\gamma$ is an introduced trade-off parameter between rate loss and reconstruction accuracy. Increasing the value of $\gamma$ will penalize more on the number of measurements and reduce the average compression ratio of $\mathcal{I}$ sets.

\section{Experiments}
In this section, we compare the proposed semantic-aware image BCS methods with fixed-ratio image BCS methods under the same number of average compression ratio. We name our method as SemBCS. The fixed-ratio version of SemBCS, FixBCS, can be obtained by using a $B\times B$ Conv with strides $B$ and $n_{avg}$ output channels as the encoder and a $1\times 1$ Conv with stride $1$ and $3B^{2}$ output channels as the decoder, followed by the same D-net used in the SemBCS. 

\subsection{Dataset}
We use two different datasets in our experiments. The first one, the MS-COCO 2014 \cite{lin2014microsoft} dataset, is composed of all kinds of images in human life and contains rich semantic information. The second one, the MPI-INF-3DHP \cite{mehta2017vnect} dataset, is widely used for human mesh recovery task and contains the video sequences where some human objects are doing some specific actions in an indoor environement with a green screen background. Therefore, the semantic information in this dataset is quite limited.

As discussed above, all images are scaled to the size of $224\times 224 \times 3$ for experiments. For the MS-COCO 2014 dataset, we use $82,783$ training samples, $2,000$ validation samples, and $2,000$ testing samples. For the MPI-INF-3DHP dataset, we first extract images from the training video sequences and then randomly choose 5\% for validation, 5\% for test, and the rest for training.

\subsection{Experimental settings} The values of $n_{b}$ and $n_{max}$ are different in the two datasets. For MS-COCO 2014 dataset, we choose $n_{b}$ from $\{150,250,350,450,550\}$, set $n_{max}=200$, and ensure $n_{avg}\approx n_{b}+100$ by tuning the value of $\gamma$. For MPI-INF-3DHP dataset, we set $n_{b}=20$, $n_{max}=200$ and ensure that $n_{avg}$ changes from around $50$ to $140$ by changing $\gamma$. The main reason for these setting difference is that the growth speed of PSNR value alongwith the increasement of compression ratio in MPI-INF-3DHP is much higher than MS-COCO 2014 dataset. In both experiments, we train the networks until the PSNR value in validation datasets stops increasing under the specific $n_{avg}$ values.

\subsection{MPI-INF-3DHP experiment}
\begin{figure}[t]
\centering
\includegraphics[scale=0.62]{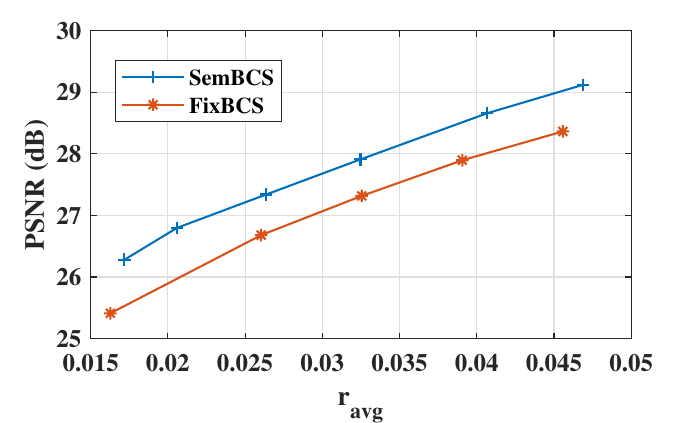}
\caption{PSNR versus average compression ratio for different methods in the MPI-INF-3DHP dataset.}
\label{fig:PSNR-MPI}
\end{figure}
We show the PSNR versus the average compression ratio $r_{avg}$ of different BCS methods in the MPI-INF-3DHP dataset in Fig. \ref{fig:PSNR-MPI}. From the figure, the SemBCS works significantly better than the FixBCS. For example, the SemBCS uses 23\% fewer samples than the FixBCS when the targeted PSNR value is 27 dB. This experiment shows the superiority of the proposed semantic-aware BCS system over traditional BCS systems.  
\begin{figure}[t]
\centering
\includegraphics[scale=0.65]{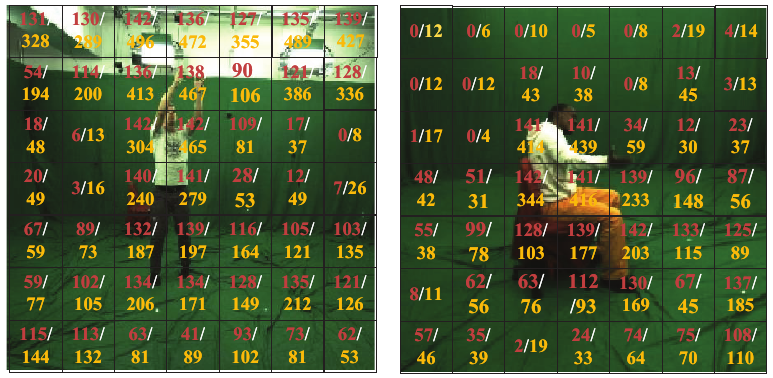}
\caption{Examples of the learned number of measurements at stage 2 (brown color) and the underlying sparsity levels (yellow color) for each block in the MPI-INF-3DHP dataset when the average compression ratio is 0.0325.}
\label{fig:EXAMPLE-MPI}
\end{figure}

To further understand the P-net learned in the SemBCS, we show some examples of the learned number of measurements at stage 2 and the underlying sparsity levels for each block in Fig. \ref{fig:EXAMPLE-MPI}. To estimate the sparsity level of each block, we solve a sparse linear inverse problem, $y=Ax$, where $y\in \mathcal{R}^{3B^{2}}$ is the vectorized version of the pixels in each block, $A\in \mathcal{R}^{3B^{2}\times 12B^{2}}$ is a predefined overcomplete discrete cosine transform (DCT) dictionary, $x$ is the underlying sparse representation. We calculate the number of elements in $x$ whose absolute value is larger than $10$ and use it as the sparsity level of each block. Note that as the D-Net is used to fuse the information from different blocks in the deep reconstruction stage, the number of measurements for one block will be affected by the neighbouring blocks. From Fig. \ref{fig:EXAMPLE-MPI}, more measurements are generally used for blocks containing richer semantic information, such as human areas, light areas, and green screens with folds. Also, more measurements are used for blocks that are less sparse, which means the P-net is able to estimate the sparsity level and assign the measurement matrix accordingly by analysing the semantic information contained in the base CS measurements. This experiment is a very good verification of our design concept.

\subsection{MS-COCO2014 experiment}
\begin{figure}[t]
\centering
\includegraphics[scale=0.65]{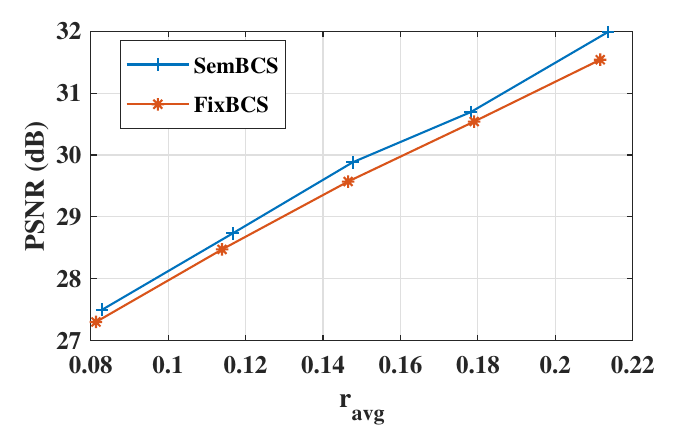}
\caption{PSNR versus average compression ratio for different methods in the MS-COCO2014 dataset.}
\label{fig:PSNR-coco}
\end{figure}
We also show the PSNR versus the average compression ratio $r_{avg}$ of different BCS methods in the MS-COCO2014 datatset in Fig. \ref{fig:PSNR-coco}. From the figure, the SemBCS still has a steady performance gain over the FixBCS, indicating the generality of the proposed semnatic-aware operations on different datasets. However, we find the performance gain is not as large as the previous experiment. This is because the P-net in SemBCS only has five Convs and is not deep enough to conduct semantic reasoning for datasets with rich semantic contents. 
\begin{figure}[t]
\centering
\includegraphics[scale=0.65]{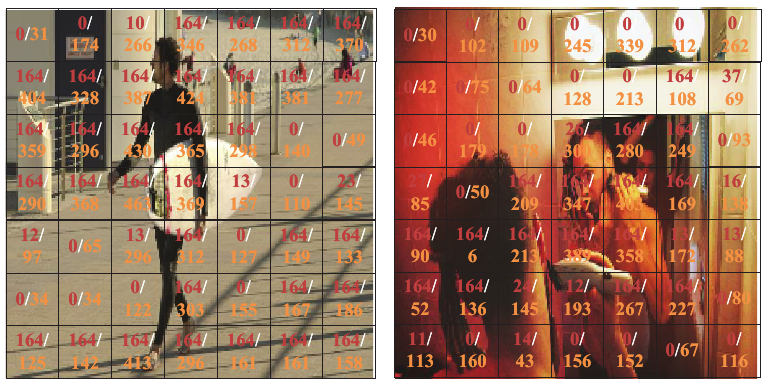}
\caption{Examples of the learned number of measurements for each block at stage 2 in the MS-COCO2014 dataset when the average compression ratio is 0.148.}
\label{fig:EXAMPLE-coco}
\end{figure}

Some examples of the learned number of measurements at stage 2 and the underlying sparsity levels for each block are shown in Fig. \ref{fig:EXAMPLE-coco}. We can see that more measurements are allocated to the human areas while fewer to the floor and the wall. However, we also notice the learned number of measurements is not strictly allocated alongwith the amount of semantic information and the sparsity levels for some blocks, which means this version of SemBCS can be further improved for datasets containing rich semantic information.

\section{Conclusion}
\label{sec:conclusion}
In this work, we have proposed a novel semantic-aware image compressive sensing system, where the best measurement matrices for different images are decided by the images' semantic information. We also verify the effectiveness of the proposed method in MPI-INF-3DHP and MS-COCO2014 datasets. Improving the architecture of the policy network and the deep reconstruction network will be left as our future work.  

\ninept
\bibliographystyle{IEEEbib}

\bibliography{strings,refs}

\end{document}